# Mechanistic Learning for Survival Prediction in NSCLC Using Routine Blood Biomarkers and Tumor Kinetics

Ruben Taieb [1], René Bruno [2,3], Pascal Chanu [4], Jin Yan Jin[5], Sébastien Benzekry [1]

1. COMPutational pharmacology and clinical Oncology Department, Inria Sophia Antipolis – Méditerranée, Cancer Research Center of Marseille, Inserm UMR1068, CNRS UMR7258, Aix Marseille University UM105, Marseille, France

2. Clinical Pharmacology, Genentech-Roche, Marseille, France
3. Current affiliation: R&B Pharmacometrics, Marseille, France

4. Clinical Pharmacology, Genentech-Roche, Lyon, France
5. Clinical Pharmacology, Genentech inc.

# ABSTRACT

**Background**

Predicting overall survival (OS) in non-small cell lung cancer (NSCLC) is essential for clinical decision-making and drug development. While tumor and blood test markers kinetics are intrinsically linked, their joint dynamics and relationship to OS remain unknown.

**Methods**

We developed a mechanistic model capturing the interplay between tumor (T) burden and three key blood markers kinetics: albumin (A), lactate dehydrogenase (L), and neutrophils (N), through coupled differential equations (termed TALN-k). This model was enhanced with a machine learning framework (TALN-kML) for OS prediction. The model was trained and validated on clinical trial data from NSCLC patients treated with atezolizumab in monotherapy (N = 862 patients) or combination therapy (N = 1,115). Model parameters were estimated using nonlinear mixed-effects modelling, and survival predictions were assessed using individual and trial level metrics.

**Results**

TALN-k successfully described individual and population-level marker kinetics, revealing complex interactions between tumor and blood markers, and improving corrected BIC and log-likelihood metrics by a significant margin of previous empirical state-of-the-art models. Feature selection methods also highlighted valuable predictive parameters, indicatives of good or poor prognosis. The TALN-kML model outperformed empirical, uncoupled models, achieving improved C-index (0.74 ± 0.02 vs 0.72 ± 0.03), 12-months AUC (0.83 ± 0.004 vs 0.79 ± 0.05), and accuracy (0.77 ± 0.03 vs 0.76 ± 0.05) in OS prediction.

**Conclusion**

Our mechanistic learning approach allows for an interpretable model, which improves on longitudinal data description and on survival prediction in NSCLC by jointly integrating tumor and blood markers kinetics. This methodology offers a promising avenue for both personalized treatment strategies and drug development optimization.

# INTRODUCTION

Predicting overall survival (OS) in oncology is critical for both personalized health care and drug development [1],[2]. For health care, patient-level predictions may guide oncologists for treatment planning and adaptation (drug, dose, sequence, scheduling). In drug development, survival predictions may help anticipating the final OS outcome of a randomized phase III trial by leveraging data from earlier-phase studies and/or interim analyses [3], [4], [5]. This is particularly critical in oncology where the attrition rate of drug candidates entering clinical testing reaches approximately 95%, including a failure rate of 40-50 % between phase III and regulatory approval [6]. While baseline variables contain predictive and prognostic information [1], incorporating on-treatment fluctuations of either tumor kinetics [7] or blood markers [8] improves the predictions. In drug development, interim analyses of phase 3 trials rely on OS surrogates based on evaluation of changes in tumor burden (i.e., progression-free survival or best overall response rates). However, their association with OS has often been found incorrect [9]. In this context, model-based parameters of the dynamics of longitudinal data carry relevant predictive value, for both individual- and trial-level predictions [8], [10], [11], [12]

Even though simple blood markers derived from routine hematology or biochemistry have been reported to be important prognostic factors of OS [13], [14], many state-of-the-art studies only incorporated baseline (BSL) values. The only longitudinal biomarker that has been extensively used and modeled to date is tumor size kinetics (TK), linked to OS by parametric survival models (TK-OS) [10], or by joint modeling with time-varying covariates [15]. TK-OS models have demonstrated strong predictive performances, but their disadvantage is to rely on imaging-based tumor size measurements, which are costly and sparse, as opposed to measures of circulating biological markers that typically occur at each drug administration.

The few studies that did investigate blood marker kinetics (BK) have mostly focused on a single marker or uncoupled, independent equations [12], [16], [17], with one notable exception [18]. Our previous work introduced a kinetics-Machine Learning (kML) approach, which showed significant improvement over TK modeling [8]. A crucial simplification was that it relied on uncoupled, empirical models for TK and each BK. By nature, these can't capture important correlations and interactions between disease progression, tumor growth and BKs. They also lack parameter interpretability, which would allow for finer tuning of the model and testing biological hypothesis. Non-trivial, complex dynamic BK profiles, either under monotherapy or combination therapy, remain to be quantitatively modeled in the context of cancer treatment, remission, and recovery.

Combining mechanistic modeling with machine learning – an approach that we named mechanistic learning [19], [20], [21] – consists in combining the strengths of each of these computational approaches. Clinical datasets with longitudinal data and overall survival typically have N ~ 1,000 patients with p ~ 50,000 longitudinal measurements. This p ≫ N (fat data) situation prevents applying straightforward machine or deep learning algorithms. Mechanistic modeling leverages biologically-based hypotheses to project patients into a lower-dimensional space of model-based parameters. Then, the dimensional setup is adequate for ML, used to agnostically learn the TK+BK – OS relationship, for which much less is known.

We propose here a novel mechanistic model of the joint kinetics between tumor burden and three blood markers, namely: albumin, lactate dehydrogenase (LDH) and neutrophils. The aim was to uncover and quantify complex and intricate interactions that hold between these BKs and tumor progression in non-small cell lung cancer (NSCLC) patients, and to extract interpretable parameters with strong prognostic values. We call this model TALN-k. In addition, we hypothesized that the incorporation of TALN-k into a ML survival model (TALN-kML) would benefit the accuracy of OS prediction. We evaluated the performances for prediction of OS, either at the patient (for clinical decision-making) or population level (for prediction of a clinical trial's outcome).

# MATERIALS AND METHODS

## Data

The data used was obtained from several clinical trials for atezolizumab (ATZ), either under combination therapy (COMBO) or monotherapy (MONO) treatment, which were analysed separately. Monotherapy data consisted of the three phase 2 clinical trials POPLAR + FIR + BIRCH (MONO, 862 patients, 44,911 total data points) [22], [23], [24]. Combination therapy data consisted of the phase 3 trial IMpower 150 (COMBO, 1,115 patients, 67,507 total data points) with 3 arms consisting of combinations of Atezolizumab (ATZ) with other agents (Bevacizumab and Carboplatin + Paclitaxel) [16].

The on-treatment variables consisted of longitudinal tumor sum of largest diameters (SLD), and blood levels of albumin, lactate dehydrogenase (LDH) and neutrophil counts. It also included baseline covariates, and time-to-event survival data.

## Data availability statement

Qualified researchers may request access to individual patient- level data through the clinical study data request platform (https://vivli.org/). For further details on how to request access, see https://www.roche.com/innovation/process/clinical-trials/data-sharing/.

## The tumor-albumin-LDH-neutrophils kinetics (TALN-k) model

Biological assumptions

The longitudinal model we developed, named TALN-k, is a joint model of coupled ordinary differential equations with delay, describing the tumor, albumin, LDH and neutrophil counts dynamics under immune-checkpoint inhibition either in monotherapy or in combination with other drugs. The model was derived assuming several pharmaco-biological hypotheses based on the literature.

Tumor dynamics modelling was linked to the data using the RECIST (response evaluation criteria in solid tumors) sum of largest diameters (SLD) of up to five target metastatic lesions. Tumor cells were assumed to be divided into two subpopulations: sensitive cells $S_1(t)$ and resistant cells $S_2(t)$. Both were assumed to have growth rate $\alpha$, but resistant cells had death rate $d_1$ and sensitive cells had death rate $d_2$. The sum of the two, $S(t)$ was the modeled version of the SLD. These assumptions lead to classical double exponential dynamics (see equations 1, 2 and 3), as first introduced by Stein et al. for prostate-specific antigen kinetics [25].

For albumin (A), LDH (L) and neutrophils (N), production was assumed to be regulated by parameters $\gamma_A$, $\gamma_L$, $\gamma_N$. In the absence of tumors, the equilibrium values of these blood markers were denoted $A_{eq}$, $L_{eq}$ and $N_{eq}$, and their clearance was assumed linear with parameters $k_A$, $k_L$, $k_N$, respectively.

Albumin production is impaired by tumor-related inflammation [26]. This mechanism was incorporated in the model by downregulating albumin production proportionally to the total tumor burden $S(t)$ with parameter $a_P$. This inflammation also increases capillary permeability as stated by Peter B. Soeters [27]. The process was modeled by modulating the elimination rate of albumin with an increasing function of $S(t)$, saturating at a value $a_I$. This yields equation 4 of our system.

Increase in LDH level may originate from two different sources. Tumor cells contain increased levels of LDH due to their switch to aerobic glycolysis for energy production, as per the Warburg effect [28], [29]. This leads to increased release of LDH upon tumor cell death, assumed to be proportional with parameter $c_0$. Another source of abnormal LDH increase follows from neighbouring stroma damage during invasion by the tumor, and the increase is proportional to the parameter $c$. This damage was only considered when the tumor grows onto the stroma. To account for this, we introduced a delay $r$ between time of tumor regrowth and the start of neighbouring tissue damage, as schematized in Figure 1. The equation obtained is (5) below.

The neutrophil model, in Equation (8), was akin to the Friberg model [30]. Precursor neutrophil cells (P) production was assumed to occur with rate $k_{PROL}$ due to haematopoiesis as in Equation (6). They further undergo a three-compartments chain of maturation (variables $T_i$) with transition rate $k_{TR}$, described by Equation (7). Precursor cell production increases with tumour-related inflammation [31]. This was modelled by an increased proliferation as a function of $S(t)$, with maximal increase $n_{INFL}$. Due to lack of PK data, the chemotherapy effect on neutrophil production was assumed constant over time. It was thus implicitly accounted for in the parameter $k_{PROL}$.

These assumptions yield the following system of coupled nonlinear differential equations (TALN-k model):

Equations

$$\frac{dS_1}{dt}=(\alpha-d_1)S_1 \quad (1)$$

$$\frac{dS_2}{dt}=(\alpha-d_2)S_2 \quad (2)$$

$$S=S_1+S_2 \quad (3)$$

$$\frac{dA}{dt}=\frac{k_A}{1+a_p S}A\left(\frac{A_{eq}}{A}\right)^{\gamma_A}-k_A\left(1+\frac{a_I S}{1+S}\right)A \quad (4)$$

$$\frac{dL}{dt}=k_L L\left(\frac{L_{eq}}{L}\right)^{\gamma_L}+c_0(d_1 S_1+d_2 S_2)+c_\square 1_{S'(t-r)>0}S'-k_L L \quad (5)$$

$$\frac{dP}{dt}=k_{PROL}P\left(1+\frac{n_{INFL}S}{1+S}\right)\left(\frac{N_{eq}}{N}\right)^{\gamma_N}-k_{TR}P \quad (6)$$

$$\frac{dT_i}{dt}=k_{TR}T_{i-1}-k_{TR}T_i \quad (7)$$

$$\frac{dN}{dt}=k_{TR}T_3-k_\circ N \quad (8)$$

Description of the population model (error model, covariates, covariance and parameter distribution)

The non-linear mixed effects (NLME) population fit was performed using the Monolix software with the SAEM algorithm. All four kinetics (tumor, albumin, LDH and neutrophils) were fit jointly, using the following error models: combined (proportional + additive), additive, proportional and proportional, respectively.

For the COMBO dataset, we used the treatment arm (between ATZ + Bevacizumab + Cisplatin + Paclitaxel, Bevacizumab + Cisplatin + Paclitaxel and ATZ + Cisplatin +

Paclitaxel) as a covariate, while the MONO dataset used none. The COMBO data and for MONO data were fitted separately.

We supposed the population parameter $\log(\theta) \sim N(\theta_{pop}, \Omega)$ to be log-normal, with a diagonal variance-covariance matrix $\Omega$.

Goodness of fit and identifiability were assessed using diagnostic plots, relative standard error (RSE) values and shrinkage of the different estimates.

BK/TK model-derived metrics

We modelled OS sequentially after TALN-K, and thus needed static metrics as parameters for the random survival forest fit. These static metrics were derived from empirical Bayes estimates (EBEs) of the longitudinal models.

Some are classical parameters, such as the tumor size relative change ($REL$) at 12 weeks, defined as $REL = \frac{S(12\,weeks)}{S(0)}$, or time to regrowth ($TTR = \frac{1}{K_S + K_G} \log\left(\frac{K_S}{K_G}\right)$), where $K_G = \alpha - d_1$ and $-K_S = \alpha - d_2$.

Others are more specific to our model: when considering the above equation on the BKs, we may interpret their kinetics as following a time-varying equilibrium state. For instance, in the absence of tumors, the equilibrium state for albumin is $A_{eq}$. But, with the presence of a tumor $S(t)$, we find that the albumin concentration will always vary in the direction of a time-varying equilibrium $A_{eq}(t) = A_{eq} \cdot \frac{1}{(1 + a_p S(t))^{\frac{1}{Y_A}} \cdot \left(1 + \frac{a_I S(t)}{1 + S(t)}\right)^{\frac{1}{Y_A}}}$. Thus, we propose that these equilibriums at 12 weeks could be informative of a patient's condition and included $A_{eq}(12\,weeks), L_{eq}(12\,weeks)$ and $N_{eq}(12\,weeks)$ in our survival analysis model.

### *The TALN-kML model for OS4*

Definition of the OS prediction problem from on-treatment data

The EBEs that were obtained using the NLME fit, along with the transformed metrics for the different kinetics and the baseline clinical covariates, were then considered as candidate predictors of OS. This two-stage approach for survival analysis, if used naively, would be erroneous because it would require to already have the knowledge of the entire on-treatment follow-up until progression, to predict OS (defined from baseline) [32], [33].

To avoid this, and to mimic a real-life clinical scenario, we placed ourselves at cycle 5 day 1 (C5D1, i.e., 12 weeks after treatment initiation). For each patient, we used only the longitudinal data available before C5D1, to identify their Empirical Bayes estimates (denoted EBEs4). We then discarded the patients who died before C5D1 and predicted OS starting from C5D1. We call this corrected prediction OS4. This way, the estimated EBEs4 may be considered already known at the time of prediction, and do not entail any of the aforementioned biases. For a visual representation of this framework and truncation, see Figure 2.

Description of kML4

The previously defined individual EBEs4 (p = 26 parameters) were added to baseline covariates (p = 10) as predictors of OS4. Following previous work [8] a random survival forest machine learning (ML) model was chosen and the complete prediction model was denoted TALN-kML4. Feature selection and feature sorting was also performed, to select informative parameters, and to see whether the BK data truly adds to the survival model. The algorithms used included LASSO, random survival forest, and univariable Cox proportional hazard methods. We then compare the individual level predictive power of four different models: TALN-k, the empirical model incorporating BK and TK proposed by Benzekry et al in [33], a model using only TK as proposed by [11], and one using only baseline variables. A 10-fold cross-validation was operated using each model, on monotherapy data and combotherapy data, separately. Goodness-of-fit metrics such as AUC, C-index and accuracy were extracted from this cross-validation and compared between models.

# RESULTS

## *TALN-k*

The TALN-k model may be seen as two main interacting components: the TK, which is the driving force of the equations, and the BK, which are impacted by the tumor properties, and add new information on the patient's overall condition. The tumor is pictured as a center piece of the model, impacting peripheral blood markers' dynamics (Figure 1A).

The tumor model, which is a simple double exponential, reflects how the cancer evolves under treatment: a sharp increase is a bad prognosis, and a sharp decrease is a good one. The time of regrowth is also a very crucial factor to predict a patient's disease evolution.

Globally, albumin dynamics are relatively straightforward: usually, albumin levels decrease when tumor size increases, and vice-versa: this may be clearly seen in individual 6, whose albumin levels increased of 15 g/L after the start of treatment. Similarly, patient number 8's relapse happens in tandem with an important drop of 20g/L, before patient death. Indeed, the albumin model, as explained earlier, follows a time-varying equilibrium state, which depends on the tumor size. If albumin concentration is below this equilibrium at time t, then the albumin concentration increases; else, it decreases. This yields a quite rich array of possible dynamics, that can look very hyperbolic in nature (as the empirical model initially proposed [8]), or can follow other non-linear, non-trivial dynamics: for instance, we sometimes see early drops in albumin that are quickly stabilized as the tumor size decreases, as for individuals 4 and 10 in Figure 3.

The LDH model may be seen as a simple controlled production model, but with two additional sources of LDH increase. The source originating from dying tumor cells not only fixes a positive correlation between tumor size and LDH levels, but also explains some early peaks of LDH in patients with strong positive response: as the treatment causes a high amount of tumor cell death, these release high amounts of LDH in return. This dynamic was seen on quite a few patient profiles, like in individuals 2, 11 and 12 of Figure 3 and the high LDH level is quickly cleared from the patient's system: this early peak is often present in patients who have a longer survival time than others.

The second source of LDH increase, from neighboring stroma damage, is only present in later stages of the treatment, after a possible patient relapse (thus, this term does not impact all patients). This very sharp increase in LDH can be seen in individuals 3, 5 and 8 of Figure 3 and usually leads to premature death shortly after: some peaks may be a sign of a positive prognostic, while others may predict the opposite.

Neutrophil dynamics were assumed to follow a very classical Friberg model, with impact on precursor cell production by the tumor. We did not incorporate a pharmacokinetic (PK) model of the chemotherapy (Cisplatin + Paclitaxel), because of lack of PK data and to avoid unnecessary complexity. In the COMBO dataset – where patients were treated with chemotherapy in addition to ATZ – the most common profile obtained with this model is an early steep drop of neutrophil count, due to chemotherapy-induced neutropenia, followed by a slower increase and/or stabilization of neutrophil count when there is remission, due to the retroactive increase of precursor cell production caused by the low levels of neutrophils. These dynamics may be seen in individuals 4,6, 10, 11 and 13 of Figure 3. In both datasets, higher and further increases in neutrophil count may be caused by remission and additional tumor regrowth, see for instance individuals 5, 7 and 9.

### TALN-k fits

First, the population fits showed very good identifiability of the model parameters, despite their large number (p = 26). Relative standard errors were found inferior to 35% on both the MONO and COMBO datasets, with only 4/40 parameters above 20% (Table 1). We also found very low η-shrinkage for all parameters, with the maximum being 9% and 11% for the separate MONO and COMBO fits, respectively. However, we also noticed high variability of some parameters, such as for r and $c_0$, which present huge parameter variances of order $10^5$%.

The model was also found to correctly fit the observed data (see the observed vs predicted and residual distribution plots in Figure 3). We also found that the corrected Bayesian Information Criterion (BIC) and the log-Likelihood scores of our model was significantly better than the one obtained with the previous empirical models[8] on both datasets, despite the much larger number of parameters. For the COMBO data, the corrected BICs and objective function values (- 2 log-likelihood) of TALN-k were respectively 5,883 and 6250 smaller than the empirical model. Similarly, TALN-k improved the BICc for the MONO data by 1,819, and the OFV by 2,142.

Another important benefit of TALN-k model is the ability to capture some erratic, non-trivial, and highly heterogeneous BK profiles, as we may see in some of the individual fits shown in Figure 4. Moreover, the semi-mechanistic nature of the model allows for interpretation of these different trends, and biological hypothesis can be drawn from them. In individual 5 of Figure 4, early peaks of LDH concentration could be captured, in tandem with a sharp tumor volume and neutrophil level decrease, due to the start of effective combotherapy tumor treatment: as tumor cells die, the inflammation is reduced, the patient feels the first effects of chemotherapy-induced neutropenia, and the dying cancer cells release LDH into the bloodstream, which could cause its transient increase. Individual 2 presents a very sharp increase of LDH and neutrophil count after a relapse in cancer progression, and followed by patient death. The LDH increase may be explained by the stroma damage term in equation (5), and suggests that severe damage to the neighbouring stroma would be the cause of the LDH peak, and would be an extremely unfavourable prognostic indicator. Meanwhile, the increase in neutrophil count may be explained by tumor-caused inflammation.

*Individual-level OS prediction*

First, random forest-based longitudinal features' importance were computed on the OS4 COMBO data. This helped to isolate covariates greatly affecting survival, which may hold medical relevance, and also to show that incorporating the BK parameters into our model actually impacts our predictions. As reported in Figure 5, the three features holding most importance were $A_{eq}(12\,weeks)$, $L_{eq}(12\,weeks)$ and the time-to-regrowth TTR. These features were specifically engineered from estimated parameters of the albumin, LDH and tumor kinetics, respectively, to be a marker summarizing the dynamics, and thus were expected to hold strong prognostic value. Noticeably, these BK-derived metrics were found more important than TK-derived ones such as $KG$, which nevertheless follows in relevance, with $k_{tr}$ and $k_{PROL}$ next, these last two being related to neutrophil dynamics. We may then infer that all BK have features that are significant in the survival prediction computed by our model.

Survival predictive metrics from 10-fold cross-validation of OS4 prediction are reported in Table 2. The C-Index, area under curve (AUC), and accuracy are consistently greater for the TALN-kML model than for others. Namely, we have the following hierarchy for all the metrics considered: TALN-kML performs better than the empirical model with TK and BK, which performs better than the model using only TK, which performs better than a model using only baseline covariates, for both MONO and COMBO. This suggests that BKs hold valuable information on the state and the future survival of a patient, and that our model captures this importance better than a fully empirical one for our OS4 prediction, which does not fall into the pit of immortal time or time-varying covariate bias.

# DISCUSSION

We introduced TALN-k, a mechanistic, organism-scale model accounting for interactions between longitudinal. When integrated into a non-linear mixed effect framework, the model was able to describe both individual-level dynamics and population-level variability, for NSCLC patients treated with an anti-PDL1, either in monotherapy or in combination. We found that this mechanistic approach outperformed non-mechanistic, empirical models [8]. Critically, despite the increased number of parameters in the model, identifiability remained excellent. The extension to prediction of OS using ML, termed TALN-kML also outperformed previous works relying on empirical models [8].

The model highlighted nonlinear patterns that warrant caution when using BK levels to assess a patient's condition. Namely, while high LDH is indeed correlated with a poor prognosis, a very efficient treatment leading to increased cancer cell death can also cause high initial peaks in LDH level, even though the tumor is shrinking. Using BK levels as static baseline covariates is not sufficient to fully and accurately capture their prognostic value.

For survival prediction, we found that TALN-kML had greater predictive power than several state-of-the-art models, such as survival models using only TK [10], [11] or using empirical, uncorrelated BK models [8]. This improvement was found in both the individual and clinical trial settings. This further reinforces the hypothesis made in [8] that there is significant prognostic value in the time evolution of BKs that is not captured by tumor kinetics and baseline variables only. Modelling BKs is specifically relevant because of their circulating nature. They can be monitored easily, frequently, and without additional burden to patients. The model could be enriched by including other BKs to TALN-k, such as C-reactive protein, other blood counts or circulating tumor DNA dynamics [34], [35]. Then, by accurately describing enough BKs, and using a fine-tuned mechanistic, correlated model we may be able to use the TK as a latent, unobserved dynamic variable in an NLME fit, thus potentially reducing the absolute need of TK data. This would allow clinicians to limit reliance on imaging to assess the total tumor burden, which is a long, expensive, and incomplete procedure. One further potential improvement could be to develop a full joint model of TALN-k and OS simultaneously, which would allow to consider the impact of dynamic metrics on survival instead of the static metrics; the fully joint modeling approach may also enable a better characterization of the variability to generate more reliable prediction intervals [15], [36].

Mechanistic modelling is an active field of research in mathematical oncology [19]. However, thus far, few models have been carefully validated with clinical data. Our study benefited from large datasets with respectively 862 and 1,115 patients, and rich sampling. This allowed to calibrate the population model and train the ML model accurately. It would be interesting to check the performances (identifiability, robustness of the OS prediction) on sparser longitudinal data, especially for tumor measurements. Further tests on data from different treatments and cancer types should also be performed to test the generalizability of the model. Namely, doing an external validation of TALN-kML on a trial not yet used for model development would also be an important stepstone to completing validation of the model.

The predictive power of TALN-kML could be useful for making individual clinical decisions, e.g., to guide second-line therapy based on TK and BK data collected during first-line treatment. Another important use would be in drug development, where GO/NO GO decisions could be informed by survival predictions of the model [3]. Indeed, decisions on whether to proceed to a following phase III when considering data from previous phase Ib,and II trials are of great importance. TALN-kML could also prove to be of great benefit in phase I trials, where TK data is sparse and limited. In a more extreme case, if we were able to model TK as a latent variable, then TALN-k could be useful during tumor agnostic phase 1 trials.

Future directions for accurately describing the intricate and complex dynamics and interactions of different BKs for a cancer patient would be to use neural ordinary differential equations, that have only been applied to TK or pharmacokinetics so far, in the field of oncology [37] To maintain some of the benefits of mechanistic modelling (namely interpretability), hybrid approaches incorporating machine-learning based equation discovery look promising [38] These are, however, recent, and should be rigorously evaluated against traditional mechanistic modelling or mechanistic learning [19], [21].

***Conclusion sentences***

TALN-k offers interpretability for combined TK-BK dynamics and outperformed previous empirical models in longitudinal data description. The OS prediction model,

TALN-kML, also exhibited improved predictive performances for immuno-monotherapy and for immuno-combotherapy.

# FIGURES

## *Figure 1: schematic + LDH stroma damage and overarching model*

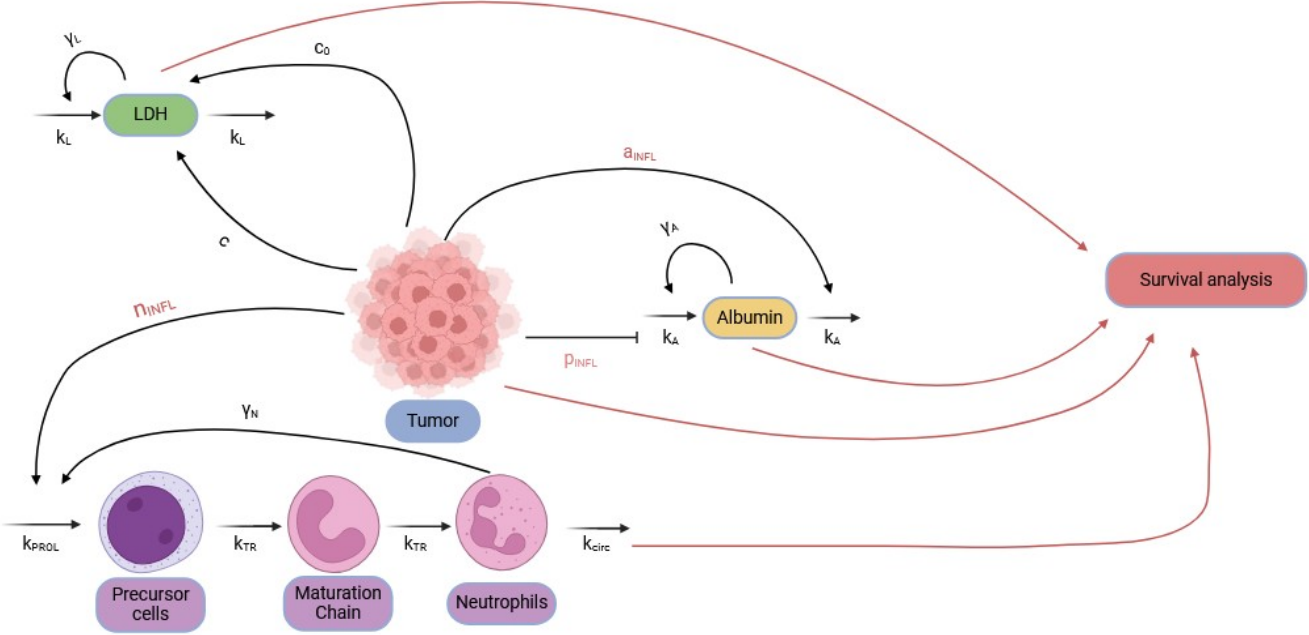

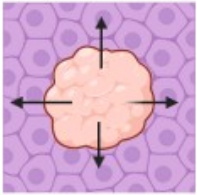
Stroma damage → LDH release

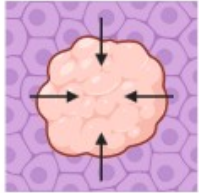
No stroma damage → No LDH release

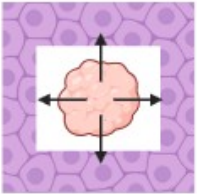
No stroma damage → No LDH release

*Figure 2: Schematic of the truncation method of OS4*

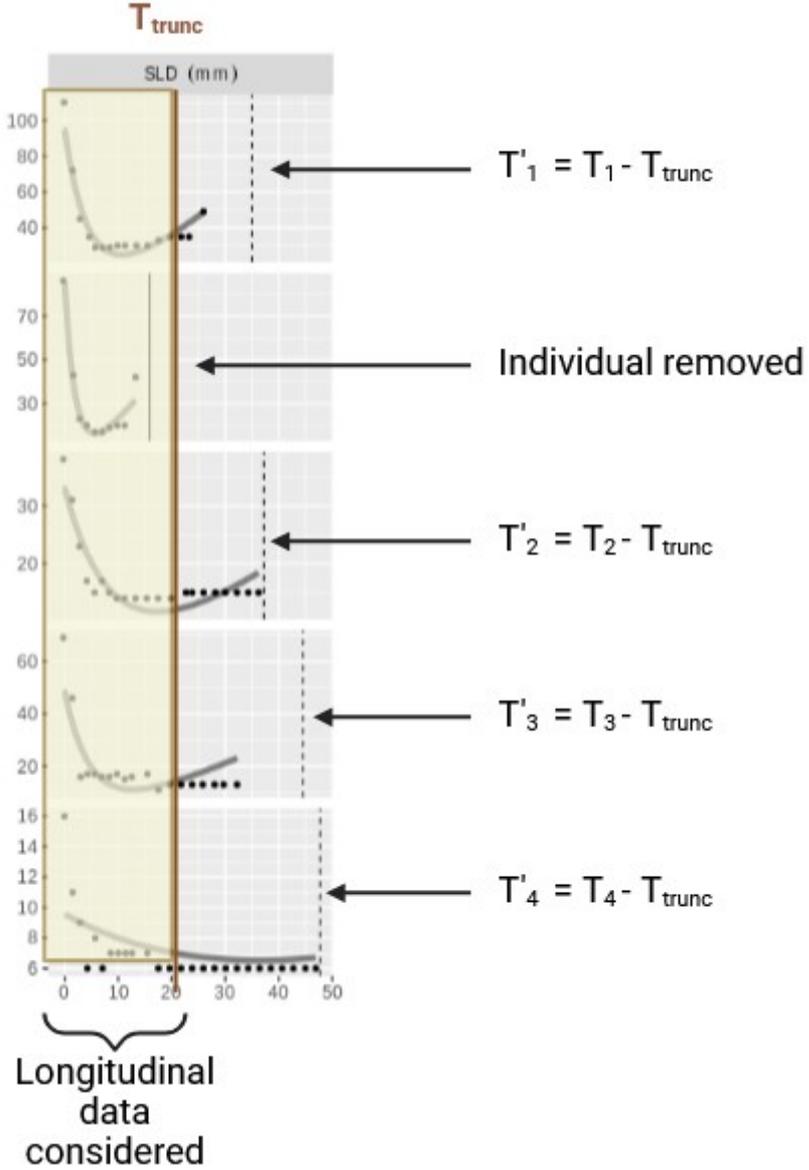

**Figure 3:** *Individual fits for TK, albumin, LDH and neutrophil count, respectively, on the combotherapy dataset. Each row represents a different individual. The vertical line can be the date of death (solid line) or censorship (dashed) of the individual.*

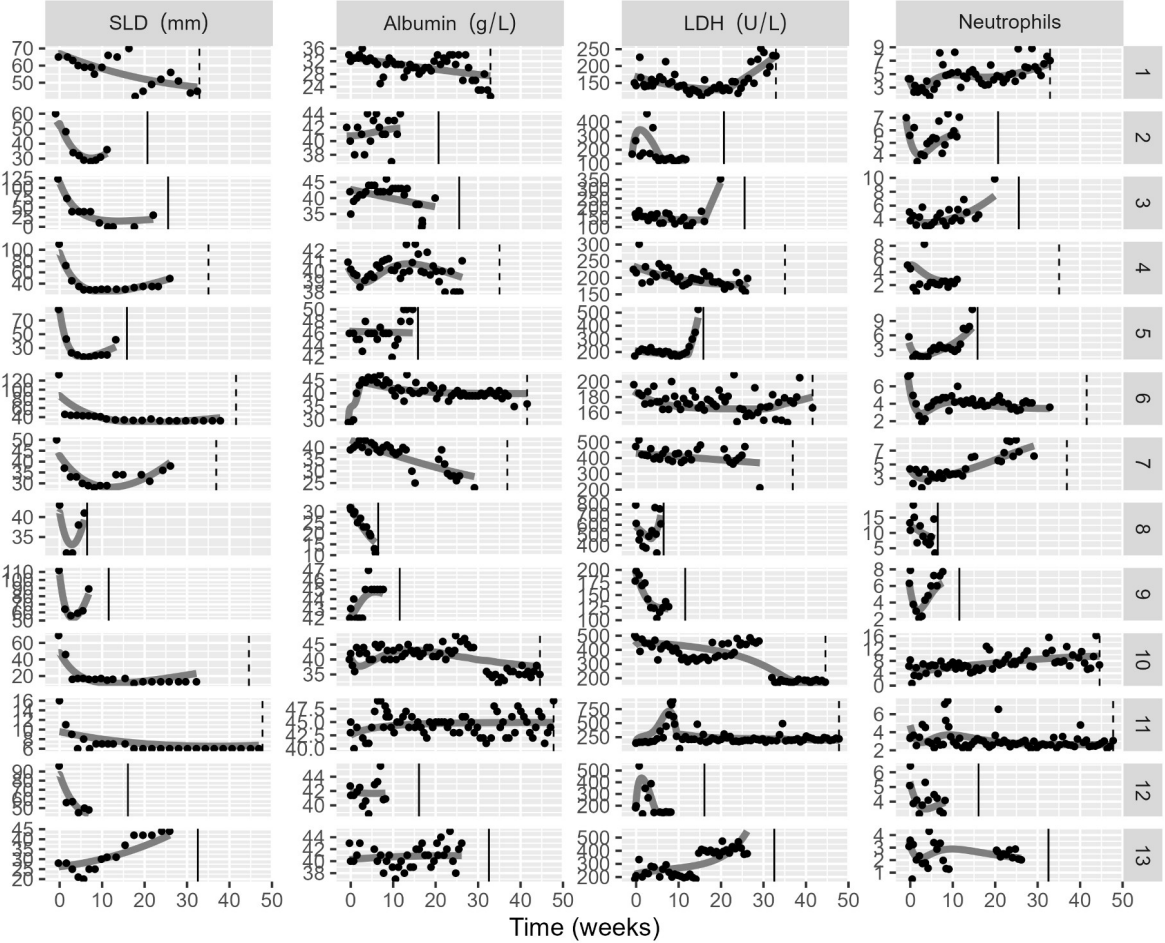

## Figure 4: Goodness of fit plots: obs vs pred + corrected VPC

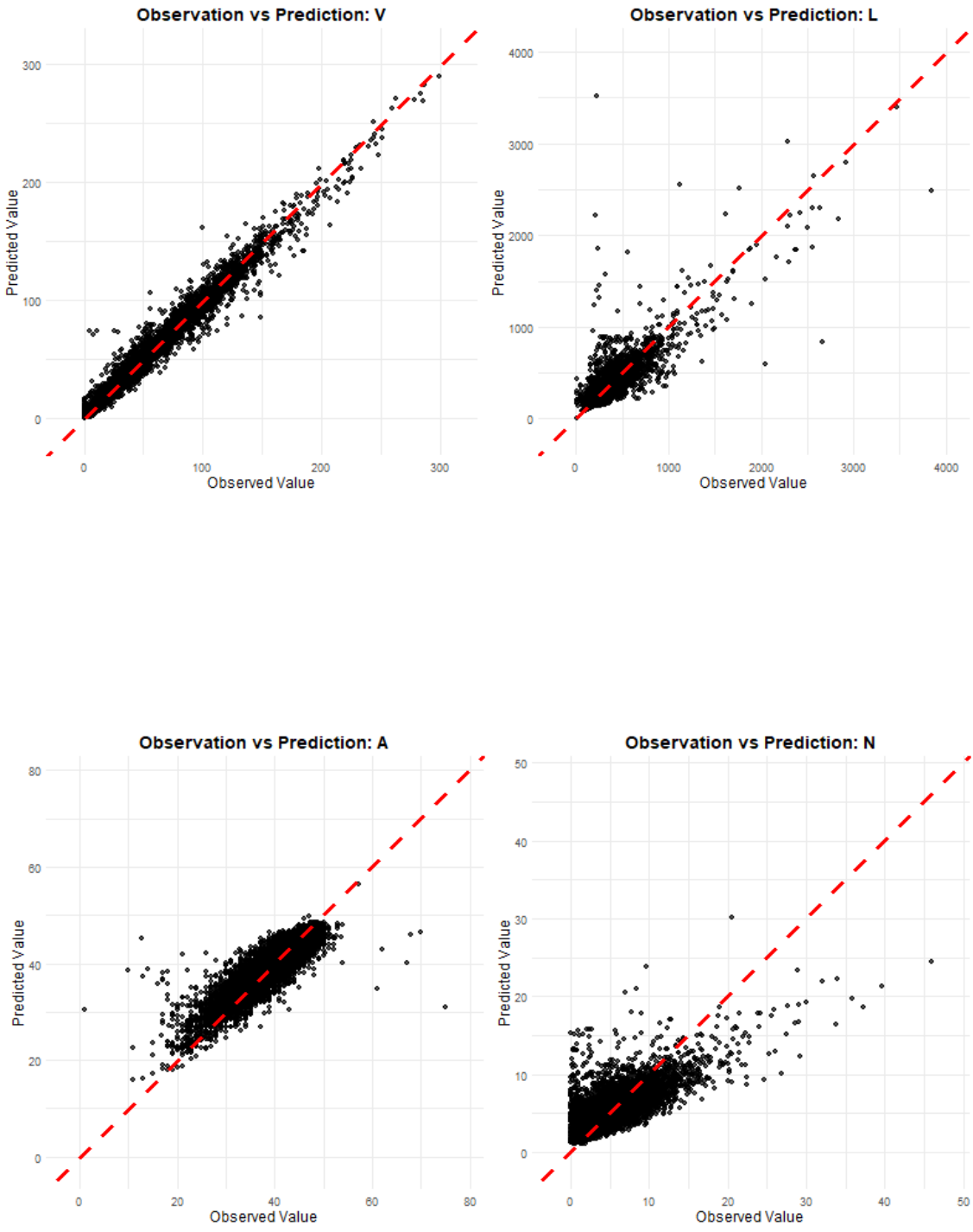

*Figure 5: Features importance*

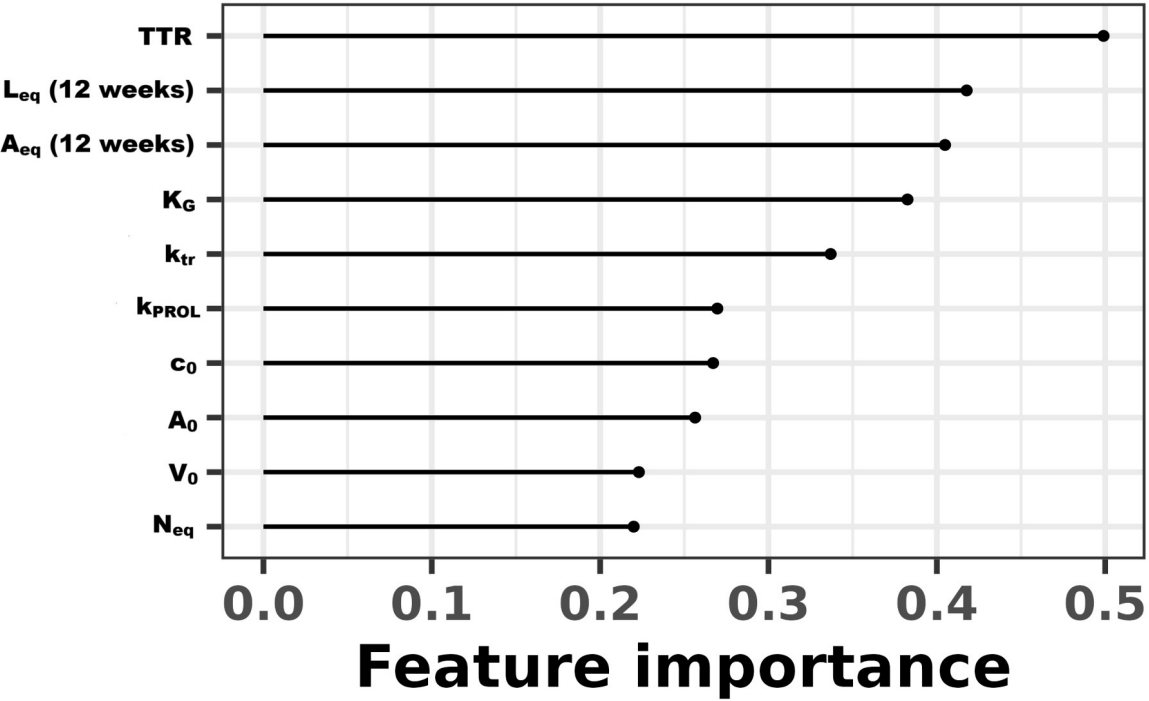

*Table 1: Model parameters*

| Parameter (unit) | Estimated Value | RSE (%) | CV (RSE %) | Meaning |
|---|---|---|---|---|
| $K_G$ (mm/day) | 0.0051 | 4.28 | 113 (2.9) | Tumor growth |
| $K_S$ (mm/day) | 0.028 | 3.89 | 108 (2.89) | Tumor shrinkage |
| $k_a$ (day$^{-1}$) | 0.00023 | 13.9 | 281 (4.65) | Albumin clearance |
| $a_p$ (mm$^{-1}$) | 11.59 | 26.6 | 418 (7.14) | Albumin production impairment |
| $\gamma_a$ (/) | 10.77 | 1.82 | 30 (4.4) | Abumin regulation |
| $A_{eq}$ (g/L) | 56.64 | 0.412 | 6 (5.87) | Albumin equilibrium level |
| r (day) | 329.31 | 21 | 253 (6.87) | Stroma damage time delay |
| $k_L$ (day$^{-1}$) | 0.0021 | 11.7 | 237 (4.81) | LDH clearance |
| $\gamma_L$ (/) | 81.02 | 1.18 | 20 (4.49) | LDH regulation |
| $L_{eq}$ (U/L) | 12.28 | 1.24 | 22 (4.34) | LDH equilibrium level |
| c (U/L) | 0.0014 | 33.8 | 384 (5.26) | LDH level in healthy cells |
| $\alpha$ (mm/day) | 0.0028 | 21.7 | 233 (5.85) | Tumor cell proliferation rate |
| $c_0$ (U/L) | 0.13 | 19.6 | 232 (5.48) | LDH level in tumor cells |
| $k_{PROL}$ (day$^{-1}$) | 0.013 | 0.86 | 12 (5.59) | Precursor cells proliferation rate |
| $k_{TR}$ (day$^{-1}$) | 0.01 | 0.633 | 9 (5.71) | Maturation rate |
| $\gamma_N$ (/) | 0.15 | 3.92 | 50 (6.05) | Neutrophil regulation |
| $N_{eq}$ (/) | 1.41 | 5.42 | 79 (5.59) | Neutrophil equilibrium level |
| $k_{circ}$ (day$^{-1}$) | 0.021 | 1.24 | 23 (4.7) | Neutrophil clearance |
| $V_0$ (mm) | 62.29 | 2.07 | 66 (2.26) | Initial tumor volume |
| $A_0$ (g/L) | 38.19 | 0.464 | 14 (2.56) | Initial albumin level |
| $L_0$ (U/L) | 259.99 | 1.62 | 50 (2.48) | Initial LDH level |
| $N_0$ (/) | 6.11 | 1.47 | 42 (2.76) | Initial neutrophil count |
| $P_0$ (/) | 7.57 | 1.97 | 28 (6.46) | Initial precursor cell count |
| $T_{10}$ (/) | 14.52 | 3.44 | 72 (4.01) | Initial count of cells in first maturation step |
| $T_{20}$ (/) | 3.56 | 2.93 | 25 (10.1) | Initial count of cells in second maturation step |
| $T_{30}$ (/) | 5.28 | 4.38 | 93 (3.83) | Initial count of cells in last maturation step |

*Table 2: Survival predictive metrics*

Courbes metrics emipirical vs mechanistic vs TK

| **COMBO** | Baseline data only | Baseline and tumor data only | Full empirical model | TALN-kML |
|---|---|---|---|---|
| C-index | 0.63± 0.04 | 0.65 ± 0.04 | 0.67 ± 0.03 | 0.68 ± 0.02 |
| AUC | 0.71±0.06 | 0.74 ± 0.04 | 0.75 ± 0.02 | 0.77 ± 0.05 |
| Accuracy | 0.69 ± 0.05 | 0.71 ± 0.03 | 0.71 ± 0.04 | 0.73 ± 0.04 |
| **MONO** | Baseline data only | Baseline and tumor data only | Full empirical model | TALN-kML |
| C-index | 0.66± 0.02 | 0.71 ± 0.03 | 0.72 ± 0.03 | 0.74 ± 0.02 |
| AUC | 0.72±0.04 | 0.78 ± 0.05 | 0.79 ± 0.05 | 0.83 ± 0.004 |
| Accuracy | 0.7 ± 0.03 | 0.74 ± 0.04 | 0.76 ± 0.05 | 0.77 ± 0.03 |

## ACKNOWLEDGEMENTS

This work was sponsored in part by Genentech. It also benefitted of funding from the French Institut National du Cancer (INCa grant #19CM148-00, ITMO Cancer AVIESAN project QUANTIC) and the French government, managed by the National Research Agency (ANR), under the France 2030 program, reference ANR-22-PESN-0017 (DIGPHAT).